\documentclass[12pt,preprint]{aastex}

\newcommand{\msun}{\, {M_{\odot}}}
\newcommand{\kms}{\, {\rm km}\, {\rm s}^{-1} }
\newcommand{\mets}{\, {\rm m}\, {\rm s}^{-1} }
\newcommand{\au}{\, {\rm AU}}
\newcommand{\etal}{et al.\ }

\shorttitle{Massive Exoplanet Inward Migration by Disk Capture}
\shortauthors{A. Font-Ribera, J. Miralda-Escud\'e, I. Ribas}

\begin{document}


\title{Protostellar Cloud Fragmentation and Inward Migration by Disk Capture as the Origin of Massive Exoplanets}


\author{Andreu Font-Ribera, Jordi Miralda-Escud\'e \altaffilmark{1} and Ignasi Ribas}
\affil{Institut de Ci\`encies de l'Espai (CSIC-IEEC), Campus UAB, Facultat de Ci\`encies, Torre C5 - parell - 2a planta, 08193 Bellaterra, Spain}
\email{font@ieec.uab.es, miralda@ieec.uab.es, iribas@ieec.uab.es}

\altaffiltext{1}{Instituci\'o Catalana de Recerca i Estudis Avan\c cats, Barcelona, Spain.}


\begin{abstract} A new model for the formation of Jovian planets is proposed. We consider planets forming at large distances from a protostar ($\gtrsim 100$ AU) through direct fragmentation of a gas cloud, by the same formation mechanism as wide stellar and brown dwarf binaries. We model the gravitational evolution of a system of these distant planets and a second population formed in a disk closer to the star. We compute the typical closest approach of these planets to the star (i.e., smallest pericenter) over the course of their evolution. When the planets reach a pericenter within a gaseous disk surroundig the star, dynamical friction from this disk slows down the planet at each plunge, causing its orbit to be gradually circularized and made coplanar with the disk. After the disk dissipates, a large fraction of these planets may be left at orbits small enough to be detected in present radial velocity surveys. A brief analytic derivation of the rate of orbital energy dissipation during these disk crossings is presented. Observational tests of this model are discussed. \end{abstract}





\section{Introduction}

Our understanding of planetary systems and brown dwarfs has been revolutionized over the last decade. About 250 exoplanets have been detected so far by the method of radial velocities, with orbital periods up to $\sim$ 10 years and radial velocity amplitudes down to $K \gtrsim 3 \mets$, although the velocity threshold for detection is still higher (K $\gtrsim 10 \mets$ ) for the majority of observed stars \citep{but06}. Several brown dwarfs have been directly imaged at large distances ($\gtrsim 30 \au$) from nearby stars \citep{nak95,giz01,neu04}, and detected as well by radial velocities especially around M dwarfs \citep{clo03}. A population of planetary mass objects has been identified in young star clusters \citep{zap00, cab07}. The majority of the exoplanets known so far are of Jovian type because of the sensitivity limit of the observations, but a few terrestrial planets have been identified by radial velocities down to masses of $\sim 5 M_E$. The microlensing technique has resulted in the detection of several additional planets of a few Earth masses \citep{bea06,gou06,gau08}, and promises to rapidly increase the detection rate in the future. Planets very close to their stars are additionally being found by transits, where a rapid increase of the number of detections is also likely from the space missions CoRoT \citep{bor03} and Kepler \citep{bas05}.

The present set of known planets and brown dwarfs already constitutes a rich data set, which informs us on the final orbital distribution of planets in semimajor axis and eccentricity as a function of the planet mass and the properties of the host star (basically mass and
metallicity). The principal challenge being faced in the field of exoplanets today is to understand how the formation mechanisms for planets and other stellar companions have determined their initial masses and orbits, and to discern how the subsequent dynamical and physical evolution of planetary systems has led to the final distribution of orbits that is observed.

Generally, two possible formation mechanisms have been discussed for Jovian exoplanets. The first is the core-accretion model \citep{ste82, pol96}, where a terrestrial planet is formed first by aggregation of planetessimals within a gas disk, and then the solid core starts accreting hydrogen and helium gas after reaching the minimum mass required for accretion to take place. The second is gravitational instability in a gaseous disk, where a Jovian planet forms directly by gravitational collapse of a gas clump in a disk \citep{bos97, bos07}. A third mechanism is the direct turbulent fragmentation of the protostellar cloud and gravitational collapse of the fragments before the gas has settled into a disk, a process that is believed to be fundamental for the formation of stars and brown dwarfs and the determination of their Initial Mass Function \citep{pad99,pad04,bon08}. This third mechanism has not been discussed as much in the literature as a possible origin for planets, with the exception of \citet{pap01} and \citet{ter02}, probably because it can only form planets at very large distances from a star ($\gtrsim 100 \au$), so it has not been thought that the planets detected by radial velocities could be formed in this way. However, \citet{raf05} has argued that the second mechanism of gas disk instability can form only massive planets at similarly large distances. In this paper, we propose a mechanism by which planets formed through the third mechanism might migrate inwards from the distances of the widest observed binary systems (probably formed by the same fragmentation mechanism) to the distances at which planets have been found in radial velocity surveys. 

A complete theory for the formation of stellar companions should be addressing at the same time the formation of planets, brown dwarfs and binary stars. It must be born in mind that the mass limits chosen to divide planets, brown dwarfs and stars into separate classes of objects (which are the minimum masses required to ignite fusion of deuterium and hydrogen, $13$ and $80$ Jupiter masses, respectively) have no relation with the physics of the formation process. We should therefore expect that, if there are several formation mechanisms, each mechanism may give rise to objects over mass ranges that partly overlap, and that extend over more than one of these three classes of objects. In fact, we already know that the core-accretion mechanism must be operating at some level, because some planets (at least Saturn, Uranus and Neptune, and the transiting planet HD149026b) have a mean density that requires the presence of a large core enriched with heavy elements. At the same time,
the presence of isolated planets in young star clusters, as well as the existence of brown dwarfs at very large distances from their primary stars (e.g., the pair of brown dwarfs at $\sim 1500 \au$ from the nearby star $\epsilon$ Indi, \citet{mcc04} favors the operation of the third mechanism as well (although some of these distant planets could be the result of dynamical ejection from a system of planets formed by core accretion). In addition, the existence of binary stars at semimajor axis $a \lesssim 30 \au$ with relatively large mass
ratios suggests that some process of gravitational instability in an opaque, disk-like structure at distances comparable to the known exoplanets can result in the formation of a massive companion by direct gravitational collapse. But could the latter two mechanisms of direct gas collapse also be responsible for some of the short-period exoplanets detected by radial velocities?

To address this question, it will be useful to list a series of relevant facts we have already learned from observations:
\begin{itemize}
\item[\em 1)] Many Jovian planets must have migrated radially, probably through their interaction with a gaseous disk, to explain their presence at orbital semimajor axes smaller than a few AU \citep{lin96}. These planets cannot form very close to their stars because the solid cores do not grow massive enough to start the process of runaway gas accretion at the high temperatures of the inner disk \citep{ste82,raf06}. In principle, this migration process could have occurred starting from distances larger than a few AU as long as there is enough gas mass in the disk at large distances. 
\item[\em 2)] The abundance of Jovian planets increases with the metallicity of the star \citep{fis05}. This must be due to a more efficient formation of planets around metal-rich stars, and not to an enrichment of the envelopes of stars hosting planets \citep{pin01}. This fact favors the core-accretion model as the mechanism for forming the majority of Jovian planets, because high metallicities would allow solid cores to grow to a larger mass and facilitate the formation of Jovian planets.
\item[\em 3)] Gas disks around young stars survive for only a few million years \citep{hai01,sic06,mor07,ale07}. Therefore, the accretion of gas by Jovian planets, as well as any migration process requiring the presence of gas, must take place during this period. Although the possibility of radial migration by planetesimals has been proposed \citep{mur98}, it seems unlikely that a disk of planetesimals could be massive enough to allow for migration of Jovian planets over large radial intervals.
\item[\em 4)] Orbits of planets have a mean eccentricity $\sim 0.3$. Their orbits have therefore been perturbed from the original circular orbits if they formed by the core accretion process, either by the interaction with the gas disk, or as a result of the dynamical interaction among several planets after the disk evaporated. The latter process may generically produce an eccentricity distribution similar to the observed one \citep{jur08,cha08}.
\item[\em 5)] There is a brown dwarf desert (i.e., a large reduction in the abundance of brown dwarfs relative to planets and stellar companions) orbiting solar type stars with orbital periods less than $\sim 10$ years. Brown dwarfs are more common at large distances, and also at short distances around M dwarf stars.
\end{itemize}

The brown dwarf desert, together with the difficulties for forming Jovian planets at small distances by gravitational instability in a gas disk (see Rafikov 2005, 2007), suggest that two mechanisms operate to form objects of different mass at short distances: core accretion, which forms planets, and gravitational instability of a massive disk or the protostellar cloud, which forms binary stars. At the same time, planets and brown dwarfs must be forming by direct gas collapse at large distances, where they seem to be abundant as well. If they were able to migrate inwards from there, a natural outcome would be that the known planet population detected by radial velocities is the result of two different formation processes, as already suggested previously (Black 1997, Mayor \etal 1998). Radial inwards migration of planets formed by core accretion must occur within a few AU, in order to explain the close-in planets, so it would not be highly surprising that similar migration ocurred at larger distance as well, owing to an extended gas disk.

In Ribas \& Miralda-Escud\'e (2007), tentative evidence was found that the host stars of the most massive planets ($M \gtrsim 4 M_J$) are less metal rich than the host stars of lower mass planets. We suggested that many of the massive planets could have formed by a process of gas collapse that is independent of metallicity, with the opacity limit for fragmentation providing a natural mass lower limit, while lower-mass planets would have formed by core accretion. We further explore this idea in this paper, focusing on the possibility of inward migration for a population of massive planets formed by gas collapse at large distances. Following the earlier suggestion of Papaloizou \& Terquem (2001) and Terquem \& Papaloizou (2002), we consider in \S \ref{sec_simu} and \S \ref{sec_resu} the dynamical evolution of a system of brown dwarfs and high-mass planets formed at large distances, together with a population of lower mass planets formed at small distances in initially circular orbits. We examine the possibility that some of the planets formed at large distances may be relocated into an orbit with a greatly reduced semimajor axis, making them detectable by current radial velocity surveys, simply by interacting gravitationally among themselves and with the population of inner planets formed by core accretion. We shall show that this dynamical evolution can easily increase the eccentricity of some objects to values close to unity, and therefore greatly reduce their pericenter, but it is much more unlikely to reduce the semimajor axis and leave the object in a stable orbit. We propose in \S \ref{sec_disk} that adding the interaction of a planet with the gas disk can act as a dissipative mechanism and help reduce the semimajor axis, achieving the desired result of migration of a planet formed far from the disk into an orbit within the gas disk at small radius. These results are discussed in \S \ref{sec_disc}.

\section{Simulations}
\label{sec_simu}

We hypothesize that there are two different populations of substellar objects orbiting normal stars. The inner population would form through gas accretion onto a rock/ice core in a protoplanetary disk, and the resulting planets would have relatively low masses (up to a few 
$M_{\rm J}$) and initially circular orbits. The outer population would comprise objects of higher mass formed by direct fragmentation of the pre-stellar cloud, in orbits of random eccentricity and inclination.

To simulate the dynamical evolution of such systems we made use of the Mercury code \citep{cha99}. After a number of initial tests employing different integrator options, we found that the most reliable results (i.e., best energy conservation characteristics) were obtained when using the conservative Bulirsch-Stoer integration \citep{pre92}. This is 
because the more sophisticated symplectic or hybrid integrators \citep{wis91,cha99} are not suitable to deal with high-eccentricity orbits, leading to fractional energy variations of a 
significant percentage. We found in our simulations quite a significant number of close encounters and resulting high-eccentricity orbits that made the choice of a conservative Bulirsch-Stoer integrator a better one because of the improved performance. In our simulations, which employed a star of one solar mass, we started with a set of initial conditions, discussed below, and then let the system evolve for 10 Myr.  We tested 
different accuracy parameters for the integrator of the Mercury code ($10^{-9}$, $10^{-12}$ and $10^{-15}$) and finally adopted a value of $10^{-12}$. This value represents the best compromise between energy conservation and computing time and results in a fractional energy change better than $10^{-4}$ in 10 Myr.

\subsection{Initial conditions of the planets}

We started with an initial inner population of 2 to 5 planets with masses randomly selected from a logarithmic distribution within 0.5 and 4 $M_{\rm J}$. The mass upper limit for planetary formation through the core accretion mechanism is not clear and it depends basically on the initial gas mass available in the disk. The upper value of 4 $M_{\rm J}$ is in agreement with the masses of heavy planets formed by core accretion in simulations in \citet{cha07}. The orbits are assumed initially coplanar and circular, and we choose the semi-major axes distributed logarithmically between 2 and 30 AU. To prevent the system from being initially unstable, we forced the planets to begin with semimajor axes with a logarithmic separation of 0.1. The rest of the orbital elements were selected randomly. Although in our Solar System there is only one planet that would fit in our population of inner objects, a number of systems have already been discovered with even three or four Jupiter-size planets in orbit within a few AU of the parent star (see the Extrasolar Planets 
Encyclopaedia for a full listing\footnote{http://exoplanet.eu}).

The outer population is also assumed to contain initially between 2 and 5 objects, with masses and semimajor axes chosen to be distributed logarithmically in the interval of 3 to 80 $M_{\rm J}$ and 30 to 1000 AU, respectively. The lower mass limit is set near the inferred minimum mass of an object formed through opacity-limited fragmentation (see \citet{cha05} and references therein for a theoretical discussion), which also implies a minimum fragmentation scale of $\sim 30$ AU, while the upper limit of the mass distribution has been selected to include objects up to the substellar limit (i.e., brown dwarfs). If the fragmentation process results in orbits randomly selected from phase space (for a fixed semi-major axis), the distribution of eccentricities and inclinations should be uniform in $e^{2}$ and in $\cos i$, respectively, and this is the assumption we made in our simulations. The rest of the orbital elements were selected randomly.

\subsection{Simulation runs}

The simulations, following the distribution of initial conditions explained above, were run as a series of 9 ensembles of 500 realizations, where the number of inner and outer planets is varied in each ensemble. This number of realizations provided sufficient data for
reliable statistics. To test the sensitivity of the results to the influence of the internal planetary system we considered two ensembles with an external population only.

Each realization was integrated for a time of 10 Myr. Typically, the evolution results in the ejection or collision of a few planets, until a small number of them are left in a quasi-stable configuration after a time that does not usually exceed 10 Myr. In tests with total integration times from 10 Myr up to 100 Myr the results showed small variations. Thus, we integrated the orbits for a time of 10 Myr in all our simulations, which is also comparable to the observed lifetime of gas disks. 

We have checked the impact of varying some of the initial conditions of the simulations on our results. For the outer population we calculated other ensembles where the lower and upper limits of the semimajor axis were changed to 50 and 500 AU respectively. We also considered wider ranges in the mass distributions of both populations (from 0.1 $M_{\rm J}$ in the inner population and up to 100 $M_{\rm J}$ in the outer). None of these ensembles showed any important qualitative differences in the results that are presented in the next section.

\section{Results}
   \label{sec_resu}

\subsection{Dynamical evolution}

A first general result we find is that the typical outcome in one of our simulated planetary systems with initially 4 to 10 objects is that a number of catastrophic events take place in which planets are either ejected to space, or collide with the central star or another planet (see Table \ref{destins}). Typically, only 1 to 3 planets out of the initial sample remain orbiting the star after 10 Myr. Most catastrophic events occur very early, within the first few Myr of evolution, before the system becomes dynamically relaxed. This fast dynamical evolution and the frequency of the various catastrophic events agree with the findings of \citet{pap01}, who considered the dynamical relaxation of a large population of giant planets similar to our outer population, as well as those of \citet{jur08}, who considered initial conditions more comparable to our inner population. Note that \citet{jur08} obtain higher rates of planet-planet collisions compared to ejections than in our case, because their planets are on average starting on smaller orbits, and the probability of planet collisions before ejection decreases with semimajor axis.

As a result of the interactions, our model generates a large number of ejected planets; over half of the initial planet population (both in inner and outer orbits) are ejected from the system. It is interesting to note that the typical ejection velocity is considerably 
lower than $1 \kms$, which is the characteristic velocity dispersion of young clusters where young planets can be observed \citep{cab07}. Therefore, most ejected planets that were born around stars within a cluster would still remain gravitationally bound to the cluster and contribute to a population of isolated planets, although some of them would escape and be observable as distant objects with proper motions consistent with their origin in an ejection event in the young cluster. The number of ejected planets is of course proportional to the initial number of planets formed on average around each star. So, clearly, if the outer population of planets is to affect the dynamical evolution of a majority of planetary systems, our model implies a total number of isolated planets that is comparable, or even larger than, the number of stars. For our chosen initial planet distribution, 2 to 6 planets are typically ejected from each planetary system.

\subsection{Semimajor axis distribution and energy transfers}

One of the inherent characteristics of the protostellar cloud fragmentation model is that planets are formed at large orbital distances. There is a minimum mass that can be reached by fragmentation determined by the opacity limit \citep{ree76}, $M_{frag} \sim 10^{-3}
\msun$. At the typical temperature of star-forming clouds, this mass is equal to the Jeans mass when the density is $10^{12} cm^{-3}$, implying a radius for a fragment of this minimum mass of $r_{frag} \sim 4 \au$. The minimum distance from a star of mass $M_{\*}$ at which this fragment may survive tidal disruption and collapse is $d_{min} = r_{frag} (M_{\*}/M_{frag})^{1/3}\sim 40 \au$. Obviously, if some of the planets discovered so far by radial velocities are to have formed by opacity-limited fragmentation, they must have been transported into a much smaller orbit than their initial one at the time of formation. A
possible way this transportation may be achieved is by a gravitational interaction with a planet formed in the disk leading to an energy exchange, in which the outer planet loses orbital energy and is transferred to a small orbit, while a less massive inner planet is
ejected. 

\begin{figure}
\centering
\includegraphics[width=8.8cm]{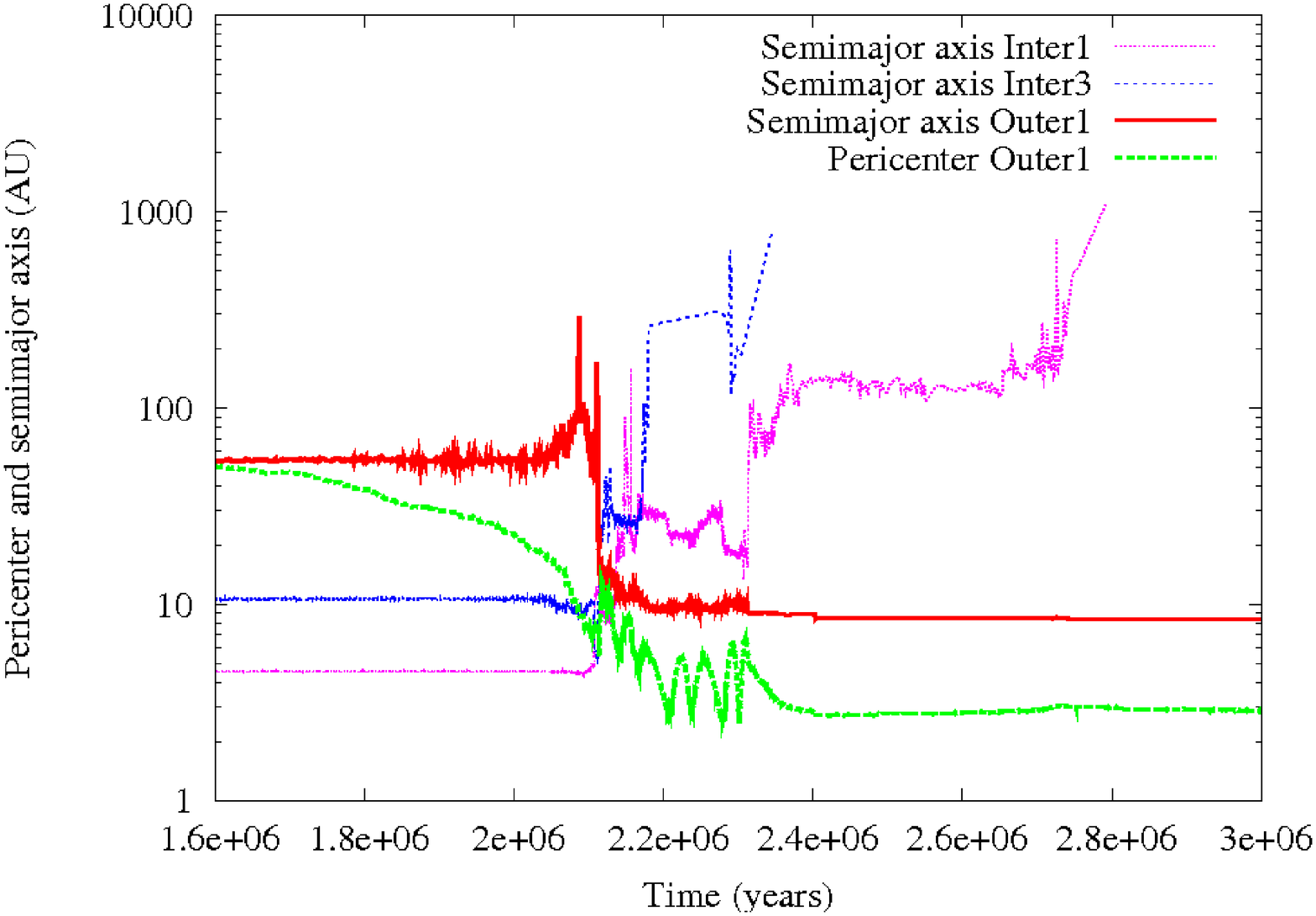}
\caption{Illustration of an orbital exchange process in one of the 
simulations. The figure shows the evolution of two of the inner planets
and a massive planet from the outer population. At time $\sim$ 2.1 Myr,
the latter planet is transferred to a small orbit after ejecting the two
inner planets.
} 
\label{exemple} 
\end{figure}

As mentioned above, our simulated systems are dynamically very active, and the migration of planets (inward and outward) is commonplace. Some of the simulations indeed show interesting cases of migration by energy transfer, as illustrated in Fig. \ref{exemple}. In this example, a planet of the outer population suffers a rapid decrease of its semi-major axis from $\sim$60 AU to $\sim$8 AU (with a pericenter distance of only $\sim$3 AU) through interation with two inner planets, which are ejected from the system.

\begin{figure}
\centering
\includegraphics[width=8.8cm]{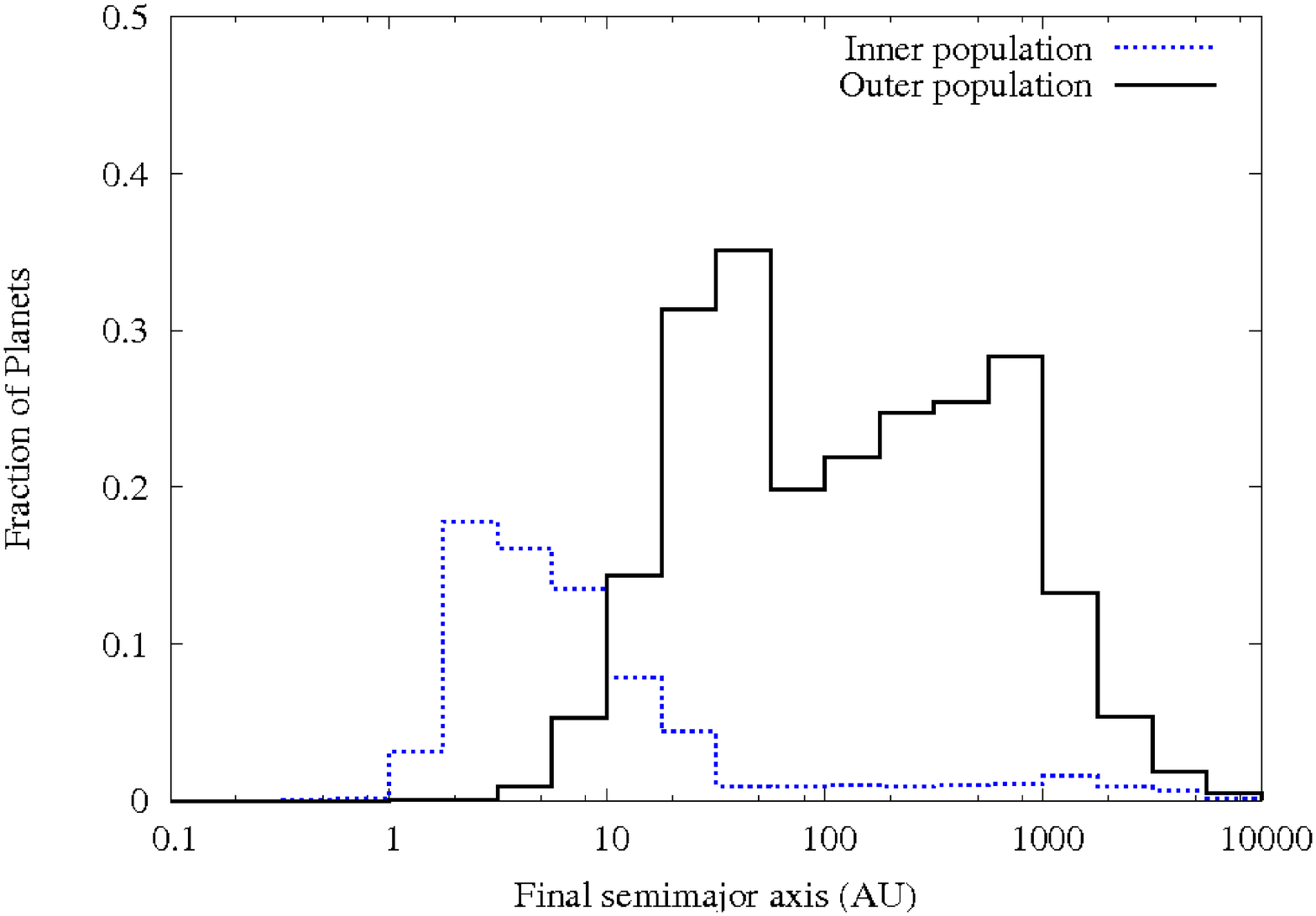}
\includegraphics[width=8.8cm]{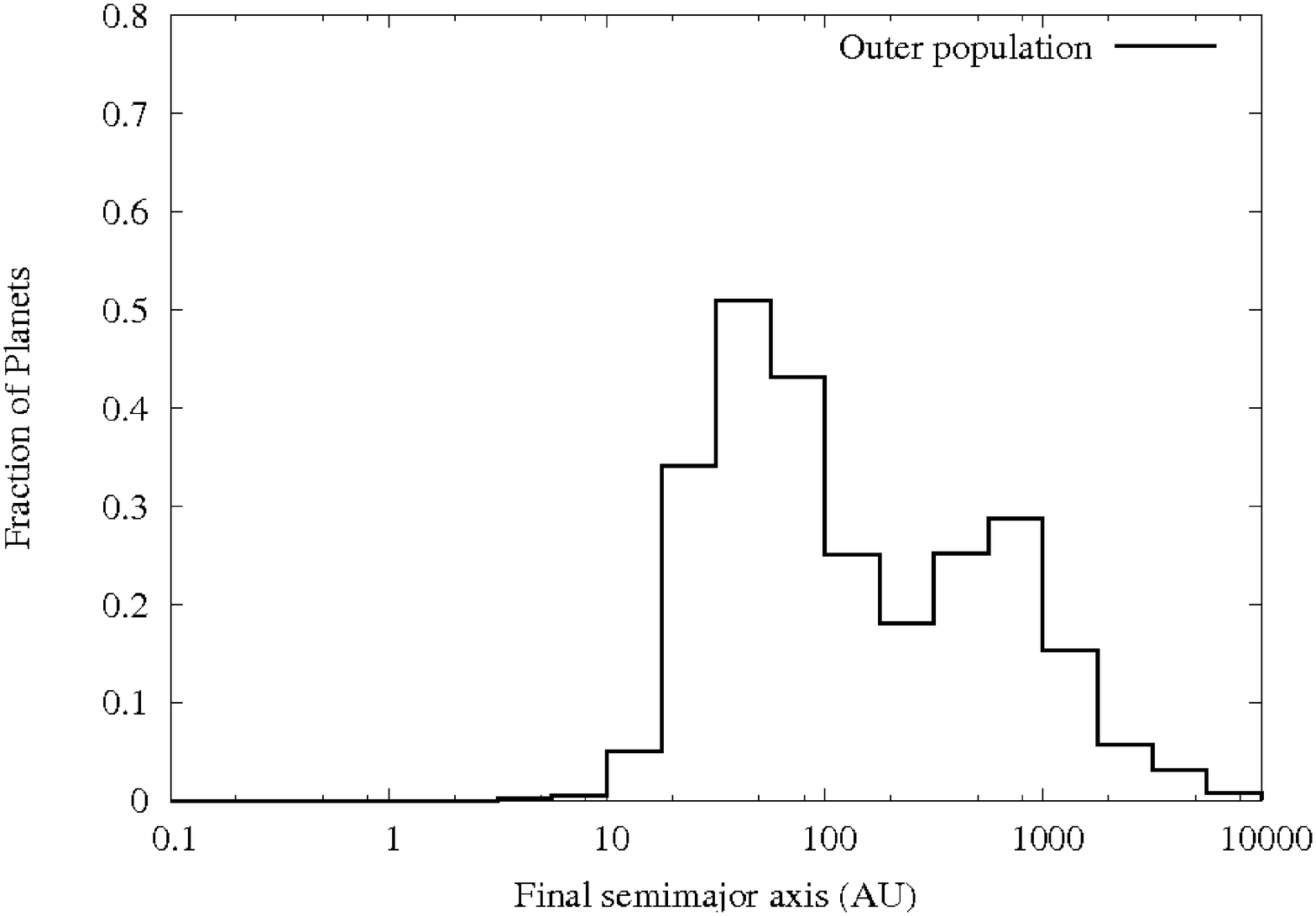}
\caption{Final semimajor axis distribution of the planet populations in 
our simulations. The left panel presents the results for the simulations 
that include both inner and outer planets, while the right panel 
corresponds to ensembles with outer planets only. The fraction of planets 
is calculated in relation to the initial planets in each respective bin. 
Note that in a log-scale axis the initial semi-major axis distribution is 
a top-hat between 30 and 1000 AU.}
\label{figeixos} 
\end{figure}

We have evaluated the frequency at which planets of the outer population end their evolution on an orbit at small semimajor axis. The results for all the ensembles of initial conditions described in \S 2.2 are shown in Table \ref{semi}. The percentage of outer planets that are
transferred to a very small semimajor axis, $a < 3 \au$, is $\sim$ 1\%, and those ending their evolution at $a < 10 \au$ are still less than 10\% of the total. The reason why changes to small orbits are very rare is simply the small phase-space volume at small semimajor axis. Even though the massive planets in the outer population tend to lose energy by giving it to lighter planets, they tend to pass close to the star only by moving to highly eccentric orbits, from which they typically end up colliding with the star or undergoing further interactions with planets with only small reductions in the semimajor axis. It is worth noticing that the percentage of outer planets with final semi-major axis below 3 AU is not affected by the presence of an internal population (note that the percentages are so small that they are strongly affected by Poisson fluctuations, since we ran only 500 simulations in each ensemble). This agrees with the findings of \citet{pap01}, who found similar rates of orbital changes in the absence of an internal population of planets. On the other hand, planets ending their evolution at $a < 10 \au$ are more clearly affected by the internal population, and in general the fraction of planets able to reduce their semimajor axis increases as the number of initial planets, both in the outer and inner population, increases.
The final distribution of semi-major axes for both populations is shown in Fig. \ref{figeixos}.

It is clear from these results that the mechanism of interactions among various planets can produce only small reductions of the semimajor axis for massive planets of the outer population, and is not sufficient to move any massive planets formed by fragmentation to the small orbits where planets have been discovered by the stellar radial velocity technique.

Even though it is improbable for planets to reduce their semimajor axis by a large factor due to gravitational interactions among themselves, the orbital evolution in chaotic systems generally leads to random changes of the orbits in which the eccentricities often reach
values close to unity. Therefore, a planet may likely have a small value of the pericenter for several orbits even if its semimajor axis does not decrease much below its initial value. If there is any additional physical process through which a planet may reduce its orbital energy (i.e., reduce its semimajor axis) when it reaches a small pericenter, the likelihood that a planet ends up in a small orbit might then be greatly enhanced. In the next section, we examine the possibility that a disk of gas or planetesimals may be responsible for
this energy loss at small pericenters. Here, we present the number of times that planets in our simulation reach small values of the pericenter during their orbital evolution.

Table 3 shows the fraction of the outer planets in our simulation that reach a pericenter smaller than $0.3$, $1$, $3$ and $10$ AU at some point in their evolution. About half of the planets reach a pericenter smaller than 10 AU. This fraction decreases rather slowly
as the value of the pericenter to be reached is decreased, and does not depend strongly on the number of initial planets in each population. Figure 3 shows the fraction of planets that reach a pericenter smaller than $q_0$ on more than the number of orbits indicated in the horizontal axis, for the case of the ensemble with 3 and 2 planets in the inner and outer populations, respectively.

As shown in the figure, a substantial fraction of planets are able to stay at these small pericenters for a large number of orbits, implying that a mechanism that reduces only a small fraction of the orbital energy at every pericenter passage may successfully bring the planet
into a small orbit.

\subsection{Final Planet Distribution}

Another interesting question that can be addressed with our simulation results is the relationship they bear on the existence of the so-called brown dwarf desert. To investigate this, we show in Fig.\ \ref{mass} the mass distribution of the planets that are left orbiting
the stars at the end of the dynamical evolution. The originally flat distribution (in logarithmic scale) is skewed towards the high-mass end after the system has relaxed. This is not a surprising result since planet-planet interactions generally affect the more massive objects the least. Therefore, massive planets can more easily avoid catastrophic events such as ejection or collision with the star. 

However, if planets are captured at different radii in the disk depending on the smallest value of the pericenter they are able to reach, then the mass distribution that should be more relevant to account for the brown dwarf desert is that of objects that reach a
minimum pericenter smaller than a characteristic value for planets detected by radial velocities. If the brown dwarf desert were a result of the dynamical evolution of the systems, then objects with mass above $\sim 15 M_{\rm J}$ would be less abundant in orbits within $\sim 5$ AU. To investigate this from our simulations, we show in Fig. \ref{nanes}
the percentage of planets at each mass bin that spend more than 10 orbits with pericenter values below a certain threshold. The results show this fraction to be nearly independent of mass. Again, this is not very surprising: planets that are perturbed into eccentricities
close to unity tend to fill the available phase space with an approximately uniform density, independently of their mass. This clearly argues against a possible dynamical origin of the brown dwarf desert arising purely from gravitational scatterings. We discuss the brown dwarf desert problem further in \S 5.

\section{Dissipation of Orbital Energy by Interaction with the Gas Disk}
   \label{sec_disk}

We have seen so far that a large fraction of planets formed at large distances may be perturbed to orbits with small pericenters by the dynamical interaction among themselves. However, a reduction of the semimajor axis takes place much less frequently. The capture of these planets into small orbits needs a dissipation process that can effectively remove their orbital energy. A possible way to dissipate the energy is by tidal interaction with the star when the pericenter is small enough, as discussed by \cite{pap01}, but this mechanism might produce only a fraction of the Jovian planets on the smallest orbits. Here, we propose another process to dissipate the energy: the gravitational interaction with a disk made of gas or planetesimals as the planet plunges through it when its pericenter has been reduced. A gaseous disk would be present in the initial stages of planet formation, and a disk of planetesimals may remain for a longer time after the gas is eliminated by accretion onto planets and the star, or by ionization and evaporation. We now do a simple calculation to
estimate the rate at which a planet may lose orbital energy by this process.

We consider a planet of mass $M_p$ plunging through a disk at radius $r$, moving along a highly eccentric orbit with semimajor axis $a$. For random orbital orientations, the radius where the planet crosses the disk will nearly always obey $r\ll a$, so the planet moves at nearly the escape speed, $v_p=(2GM_{*}/r)^{1/2}$, where $M_{*}$ is the mass of the
star. We assume that this velocity $v_p$ is much smaller than the escape speed from the surface of the planet, which is correct in most cases of interest (e.g., a gas disk may extend out to $\sim 20 \au$, where the escape speed is $\sim 10 \kms$, whereas a typical escape speed from a Jovian planet is $\sim 60 \kms (M/M_J)^{1/2}$; this escape speed is
lower for young planets of age 1 Myr by a factor $\sim 1.5$ only, see Burrows \etal 1997). In this case, the change in momentum given to the planet as it crosses the disk is dominated by the gravitational perturbation the planet induces on the disk material, which then slows
down the planet gravitationally (essentially, a dynamical friction effect).

For simplicity, we consider only the case when the planet crosses the disk at a perpendicular angle, in which case the crossing radius $r$ is twice the pericenter (for eccentricities close to 1). In reality one should consider all possible orbits at any random angles of the inclination and longitude of the periastron to find the average capture rate of planets, but the example of an orbit crossing the disk perpendicularly at pericenter will suffice for the simple estimate we wish to make here. In practice, the capture rate should be faster for most other orbits. The disk material is moving on a circular orbit at velocity $v_p/\sqrt{2}$, and therefore the relative velocity between the planet and the disk fluid is $v_r = \sqrt{3/2} v_p$.

A fluid element in the disk at an impact parameter $b$ from the trajectory of the planet will experience a velocity deflection 
\begin{equation}
 v_{\bot} \simeq \frac{2GM_p}{v_r b} =  \frac{v_p^2}{v_r} \frac{b_1}{b} ~.
\end{equation}
Here, $b_1 \equiv r M_p /M_*$ is the minimum impact parameter where the impulse approximation is valid. Using energy conservation, the planet is slowed down by the disk material deflected behind it, reducing its velocity along its direction of motion by
\begin{equation}
\Delta v \sim \frac{v_{\bot}^2}{2v_r} \frac{\Delta m}{M_p} ~,
\end{equation}
where $\Delta m$ is the mass of the fluid element in the disk that is deflected. Hence, dividing the disk into annuli at different impact parameters $b$ with mass $\Delta m \simeq \Sigma (r) 2 \pi b \Delta b$ (neglecting the variation of $\Sigma(r)$ around the annulus), the
total variation in the velocity of the planet is:
\begin{equation}
\Delta v \simeq \frac{m_1}{M_p}v_r + \int_{b_1}^r db\, \frac{v_{\bot}^2}{2v_r}
\frac{2 \pi b \Sigma (r)}{M_p} \\
\simeq \frac{m_1}{M_p}v_r \left[ 1 + \left( \frac{v_p}{v_r}\right)^4
\ln \left(\frac{M_*}{M_p}\right) \right] ~.
\end{equation}
We have added the contribution to $\Delta v$ from the fluid at $b < b_1$, approximated as $\Delta v \sim v_r m_1/M_p $, where the disk mass inside $b_1$ is $m_1 = \pi b_1^2 \Sigma(r)$.

  Assuming the following model for the surface density of the disk:
\begin{equation}
  \Sigma(r) = {M_d \over \pi R_d^2}\, \left( {R_d\over r} - 1 \right) ~,
\end{equation}
where $M_d$ is the total mass of the gas disk and $R_d$ is an outer cutoff radius, we obtain the following expression for the planet velocity change each time it plunges through the disk:
\begin{equation}
{\Delta v \over v_p} \sim  \frac{M_p M_d}{M_*^2} \left( \frac{r}{R_d}\right)^2 \left(\frac{R_d}{r} -1 \right) \frac{v_r}{v_p} \left(1 + \left(\frac{v_p}{v_r}\right)^4 \ln \left(\frac{M_*}{M_p}\right) \right)
\label{dvp}
\end{equation}

If the planet starts with an orbital semimajor axis $a$, we can consider that it has effectively been captured into the disk once the semimajor axis has appreciably decreased owing to the energy dissipation during disk crossings, even though the velocity at pericenter will have decreased only by a small amount if the orbit is highly eccentric. The reason is that the orbital period will have substantially decreased and, if the pericenter has remained approximately the same, the planet will continue to plunge through the disk until its orbit is made coplanar with the disk and is circularized. A planet can only escape this fate if it is perturbed by other distant planets to a larger pericenter before dissipating its orbital energy, which is likely to happen only before the semimajor axis has decreased substantially.

Since the orbital energy of the planet is $GM_{*}M_p/(2a)$, and the energy dissipated during a disk crossing is $M_p v_p\, \Delta v_p$, the number of disk crossings required for capturing a planet into the disk is 
\begin{equation}
 N_{dc} = {GM_{*} \over 2a\, v_p\, \Delta v_p } ~.
\end{equation}
Substituting $v_p=(2GM_{*}/r)^{1/2}$, and using equation (\ref{dvp}), we obtain
\begin{equation}
N_{dc} = \frac{R_d^2}{4r \, a} \frac{M_{*}^2}{M_p \, M_d} \frac{v_p}{v_r}  \left(\frac{R_d}{r}-1\right)^{-1} 
\left(1 + \left(\frac{v_p}{v_r}\right)^4 \ln \left(\frac{M_{*}}{M_p}\right) \right)^{-1}
\label{ncross}
\end{equation}
Note that when the disk surface density profile is $\Sigma(r) \propto r^{-1}$, the number of disk crossings required to slow down the planet at fixed semimajor axis is independent of $r$ for $r<<R_d$. When the crossing radius decreases, the higher disk surface density and higher
planet velocity increase the energy dissipated per unit of perturbed gas mass, but the impact parameter $b_1$ decreases and therefore a smaller gas mass is perturbed by the planet.

In order to check the validity of this analytical derivation, we carried out the following simple numerical test: we emulated a disk with $N$ equal mass planetessimals in circular and coplanar orbits, distributed in semimajor axis to reproduce the disk density profile, and followed the evolution of a massive planet crossing the disk with a perpendicular orbit to obtain the fractional energy variation per orbit. Using a large enough number of planetessimals ($N\gtrsim 100$) to reduce the dispersion and averaging over a large number of simulations to reduce the statistical error, we found that equation \ref{dvp} works properly (within a factor 2) over a wide range of parameters.

Considering now a typical case of a planet with $M_p/M_{*} = 10^{-2}$, and a disk with mass $M_d/M_{*} = 10^{-1}$ and radius $R_d = 100 \au$, we find from equation (\ref{ncross}) that if the planet starts on an orbit with $a= 1000 \au$ and crosses the disk at a radius $r<<R_d$, the number of crossings required for capture is $\sim 7$. Multiplying by the
orbital period for $M_{*} = \msun$, the time required for capture is less than $\sim 10^5$ years. Hence, we see that even for a disk with the mass of the minimum solar nebula ($\sim 0.04 \msun$), most planets formed at large distances that are perturbed to a pericenter where they intersect the gas disk would be captured into the disk within the observed lifetimes of disks around young stars of a few million years. In fact, this process of planet capture seems inevitable, in view of the observations we already have of the presence of gas disks 
and of brown dwarfs and massive planets at large distances.

In general, if massive planets formed by fragmentation at large distances undergo random orbital perturbations, they will uniformly fill the phase space region of small pericenters where they cross the disk. This means that the distribution of their pericenters, $q=a(1-e)$, will be uniform at small $q$. For a surface density profile $\Sigma (r) \propto r^{-1}$, the planets will also be removed from these orbits and brought inside the gas disk at a uniform rate, so they will be inserted in the disk with a distribution of orbital radii that is uniform in $r$. Most planets would therefore be placed at large radius, but a fraction of them would directly be left within a few AU, where they can be detected in radial velocity surveys. In general, if $\Sigma(r) \propto r^{-\alpha}$, then the number of crossings required for capture is $N_{dc} \propto \Sigma(r) r \propto r^{1-\alpha}$, so steeper disk profiles would lead to a greater fraction of planets being captured at small radii in the disk. However, if relaxation of the distant planet orbits is slow, an ``empty loss-cone'' distribution of the planet orbits is produced, and planets should start interacting with the gas disk at large radius as they undergo small perturbations at every orbit, so they could more often be placed at large radius in the disk. Planets may of course also migrate radially after they are captured. Therefore, their final distribution in semimajor axis likely depends on many complicating factors.

\section{Discussion}
\label{sec_disc}

We propose in this paper a new model for the formation of the most massive Jovian planets based on direct fragmentation of a pre-stellar cloud. This fragmentation process would form planets at distances from the star much larger than the orbital sizes that are detectable in radial velocity surveys, implying that a migration mechanism is required to transport these planets to orbits with semimajor axis much smaller than their initial values at formation. Using orbital evolution simulations we have explored the orbital exchange mechanism, in which a massive planet formed by fragmentation would gradually be transported to a small orbit by gravitational interactions with other smaller planets in the system. The results show that only about 1\% of the planets born beyond 30~AU reach final states with semi-major axes smaller than about 3 AU.

A better mechanism to transport planets to small orbits is obtained when considering the dynamical friction effect with a disk around the star. In this case, a planet can be transported inwards by decreasing only the pericenter through perturbations with other planets, therefore increasing by a large factor the amount of phase space from which the
required migration can be achieved. Our results show that a significant fraction of the outer planets spend several orbits at pericenter distances below 3 AU over their dynamical evolution. The approximate analytical estimate in the previous section confirms that the semimajor axis can be reduced with a relatively small number of disk crossings, placing a planet formed by fragmentation into an orbit of the typical size detected by radial velocities.

This dynamical mechanism for migration does not provide an explanation for the absence of brown dwarfs in orbits with $a \lesssim 5$ AU around solar-type stars (the so-called brown dwarf desert). As explained in \S \ref{sec_resu}, the probability for a planet to reach a small pericenter leading to disk crossings is nearly independent of its mass. In addition, in \S \ref{sec_disk} we showed that the number of disk crossings needed to slow down a planet is inversely proportional to the mass (see eq. \ref{ncross}), suggesting that brown dwarfs should
actually be easier to capture than Jovian planets. However, a possible difficulty for capturing objects that are very massive may be that the disk quickly destroyed by the plunges themselves as a result of the induced heating. Brown dwarfs might simply be too massive to undergo the process of orbital grinding before they destroy the gas disk. If it is captured at large radius in a gas disk, a brown dwarf might also not be able to migrate inwards because its angular momentum would easily be larger than that of the disk.

The scenario of direct fragmentation we have proposed may be tested observationally in several ways, and we discuss some of them here: 
${\emph i)}$ As explained in \S \ref{sec_resu}, the dynamical evolution of planetary systems that perturbs planets into the high eccentricities required for interaction with the disk should also produce a large number of ejected planets. If the star hosting the planetary system is part of stellar cluster, the ejection velocity of the planet would often be small enough for the ejected planet to remain gravitationally bound to the cluster, implying
that the abundance of isolated planetary-mass objects should be comparable to that of stars. Some isolated planets have in fact been identified in young star-forming regions \citep{zap00,cab07}, and our hypothesis may be tested as the observational determination of the abundances of these objects improves. Note, however, that many planetary-mass objects
may be also be born by fragmentation without being initially bound to any star, so the abundance of isolated planets bound in clusters can only yield an upper limit to the number of planets that are initially bound to an individual star. Some of the ejected planets will have a high enough velocity to escape from the cluster, and these may in principle provide a more specific test for our model: they should be discovered around young open clusters and associations moving away from the clusters at the typical ejection velocities ($\sim 1 \kms$),
with a distribution of distances that would indicate the rate at which they have been ejected from the young planetary systems.

${\emph ii)}$ If massive planets are indeed formed by gas fragmentation, they should probably not contain a rocky core (although see \citet{bos02} for the possibility that a core could form by sedimentation of dust grains to the center), in which case they would be less dense than objects formed by core accretion. It might be possible to test this for planets of known age with measured radii from transits. These will probably have to be, however, transiting planets that are not very close to their host stars, to avoid possible effects from gravitational tides and the stellar radiation that seem to be affecting the radius evolution \citep{bod01,gui06}.

${\emph iii)}$ Formation of planets in a protoplanetary disk will naturally predict that the spin axis of the star and the orbital axis of the planet are in close alignment. However, formation by fragmentation in a random pre-stellar cloud would result in a random orientation of the planet orbit relative to the star spin. When the planet is captured by the disk, the disk plane will be changed for a massive enough planet, when the planet mass is comparable to that of the disk, in which case a large misalignment of the star spin and the planet orbit would be expected. This may be tested for transiting planets, for which the stellar spin orientation can be measured from the Rossiter-McLaughlin effect \citep{que00,win05}.
Interestingly, the most massive transiting planets, such as XO-3 b (11.8~M$_{\rm J}$) or HD 17156 b (3.1~M$_{\rm J}$), do indeed tend show a moderate degree of misalignment \citep{nar08,heb08}.

\acknowledgments

We would to acknowledge stimulating discussions with John Chambers, Mario Juric, and Man Hoi Lee. The authors acknowledge support from the Spanish Ministerio de Educaci\'on y Ciencia via grant AYA2006-15623-C02-01. A. F.-R. thanks the Spanish CSIC for support via a JAE-PreDoc research fellowship.

\clearpage


\begin{figure}
\centering
\includegraphics[width=8.8cm]{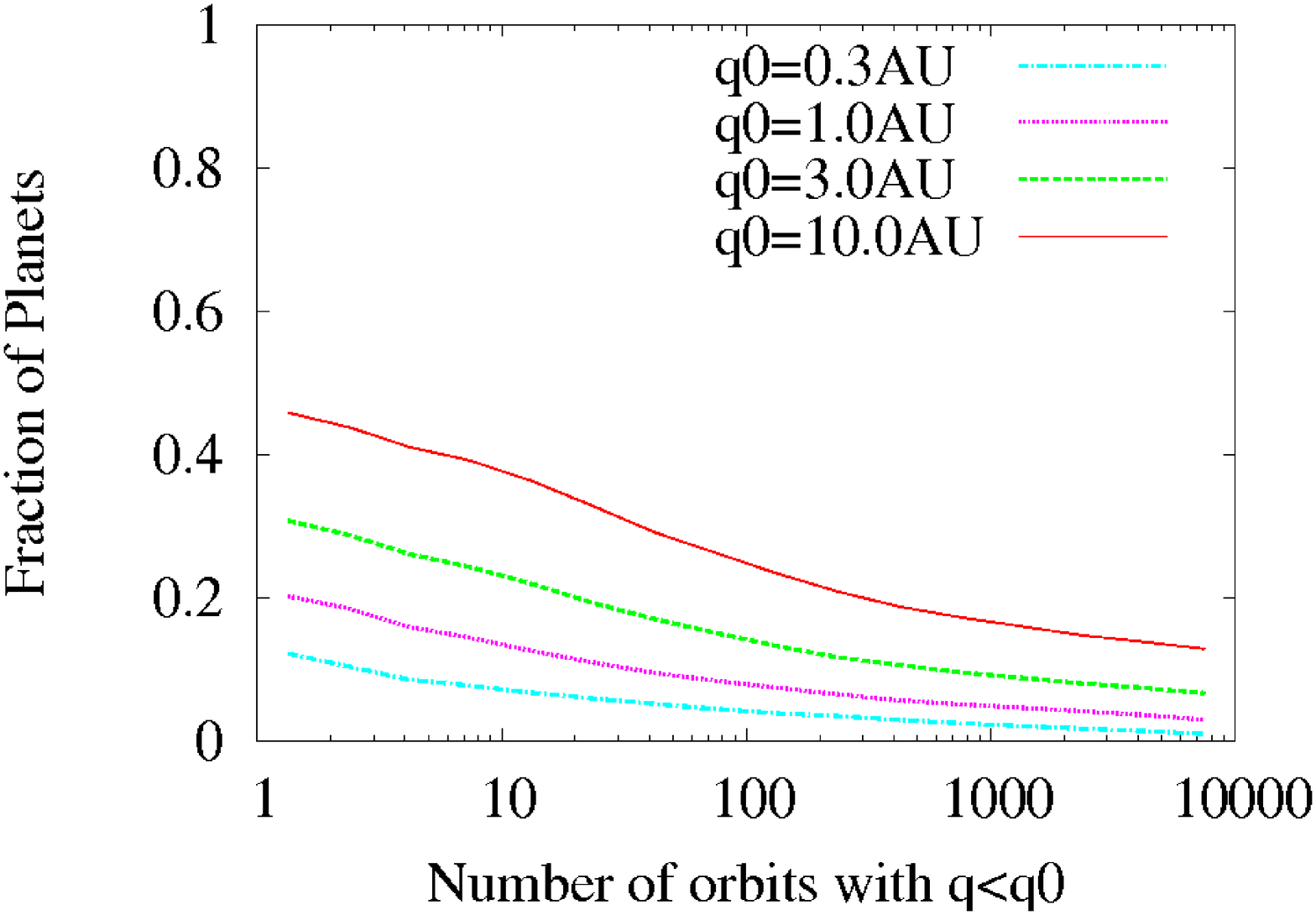}
\caption{Fraction of planets that stay a certain number of orbits (the 
abscissa value) with a pericenter lower than $q_{\circ}$.}
\label{figq0}
\end{figure}

\begin{figure}
\centering
\includegraphics[width=8.8cm]{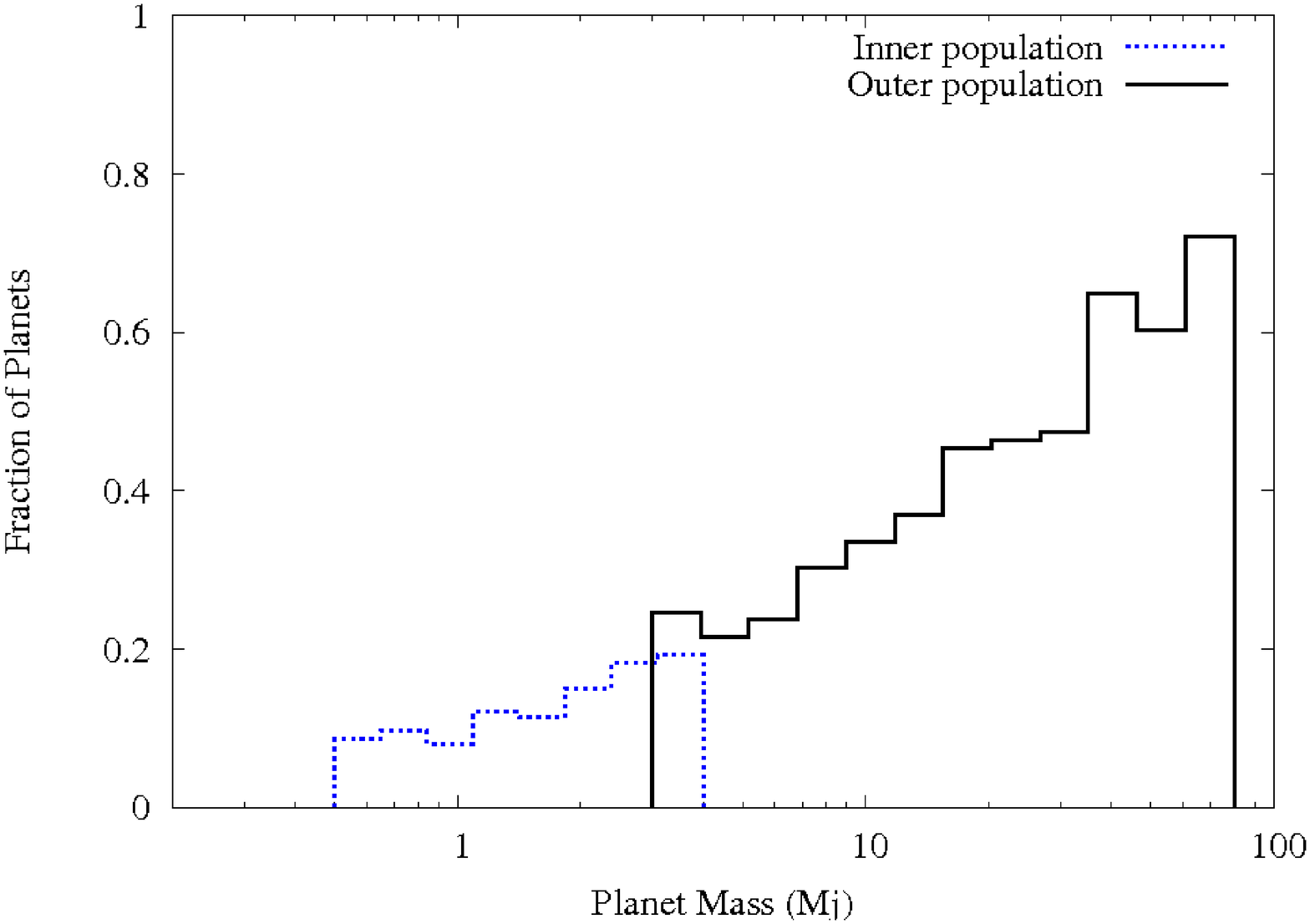}
\includegraphics[width=8.8cm]{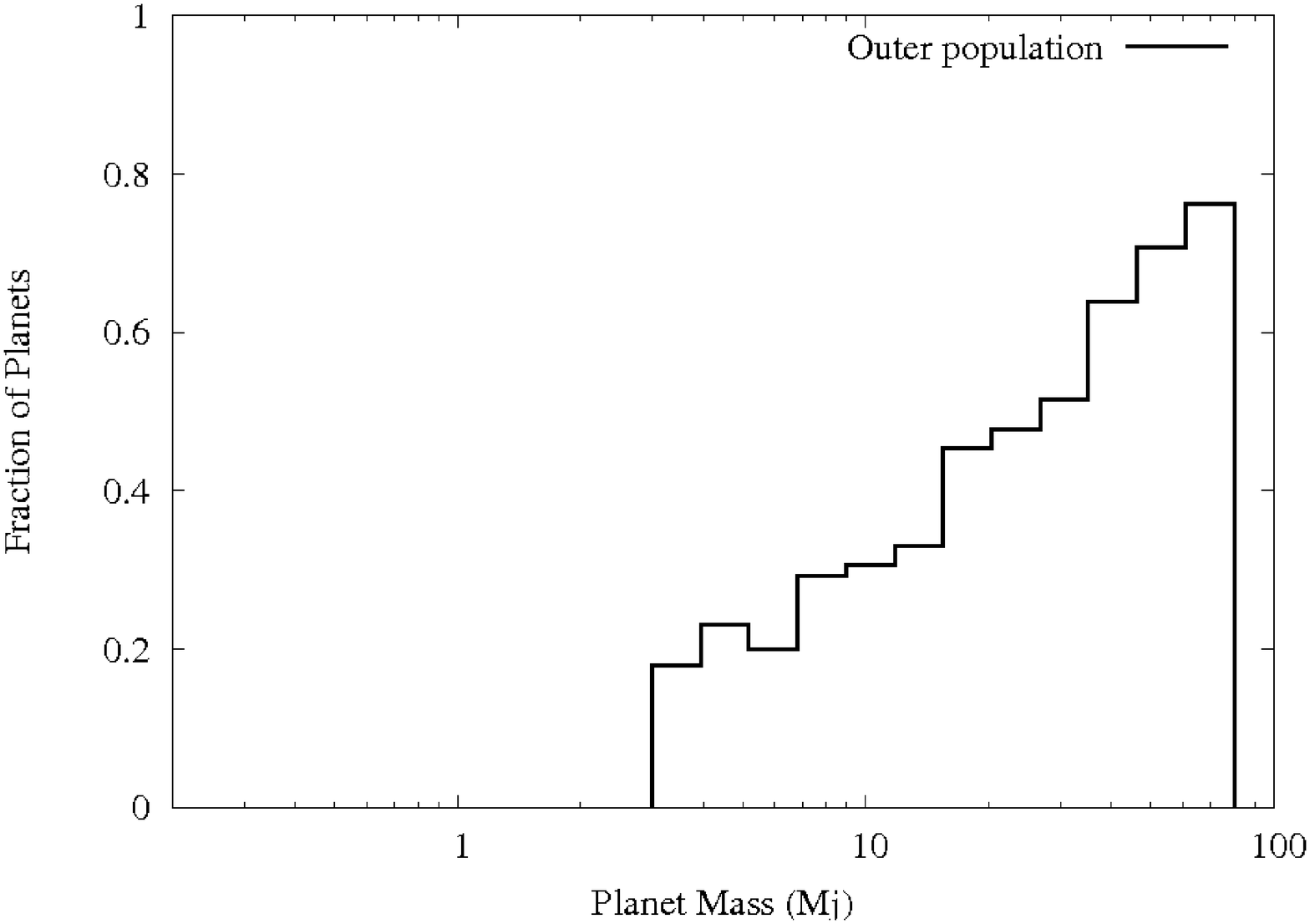}
   \caption{Final mass distribution of the planet populations in our 
simulations. The top panel presents the results for the simulations that 
include both inner and outer planets, while the bottom panel corresponds 
to ensembles with outer planets only. The fraction of planets is 
calculated relative to the initial number of planets in each respective 
bin. Note that the distribution was initially flat in logarithmic scale.}
\label{mass} 
\end{figure}

\begin{figure}
\centering
\includegraphics[width=8.8cm]{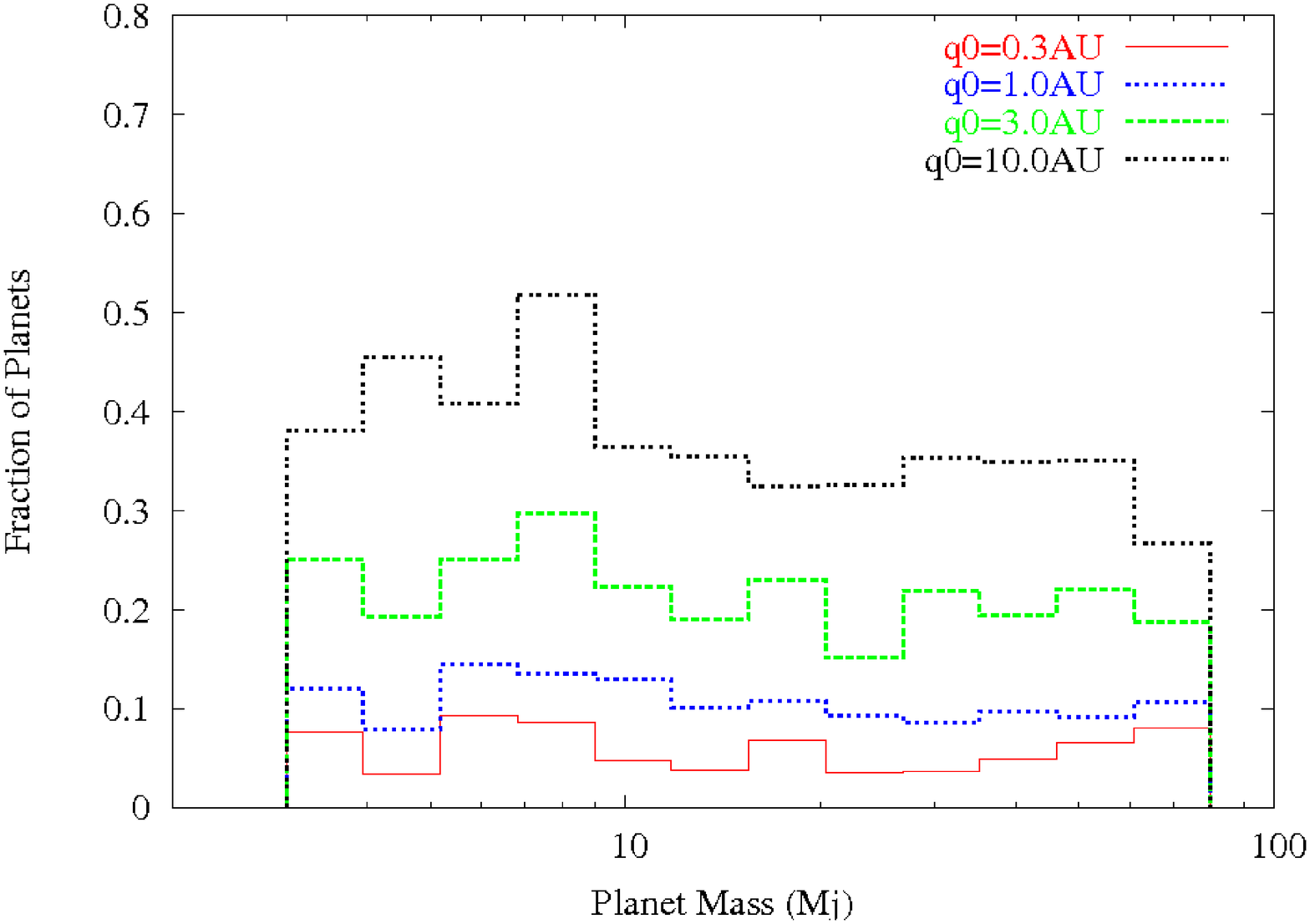}
\includegraphics[width=8.8cm]{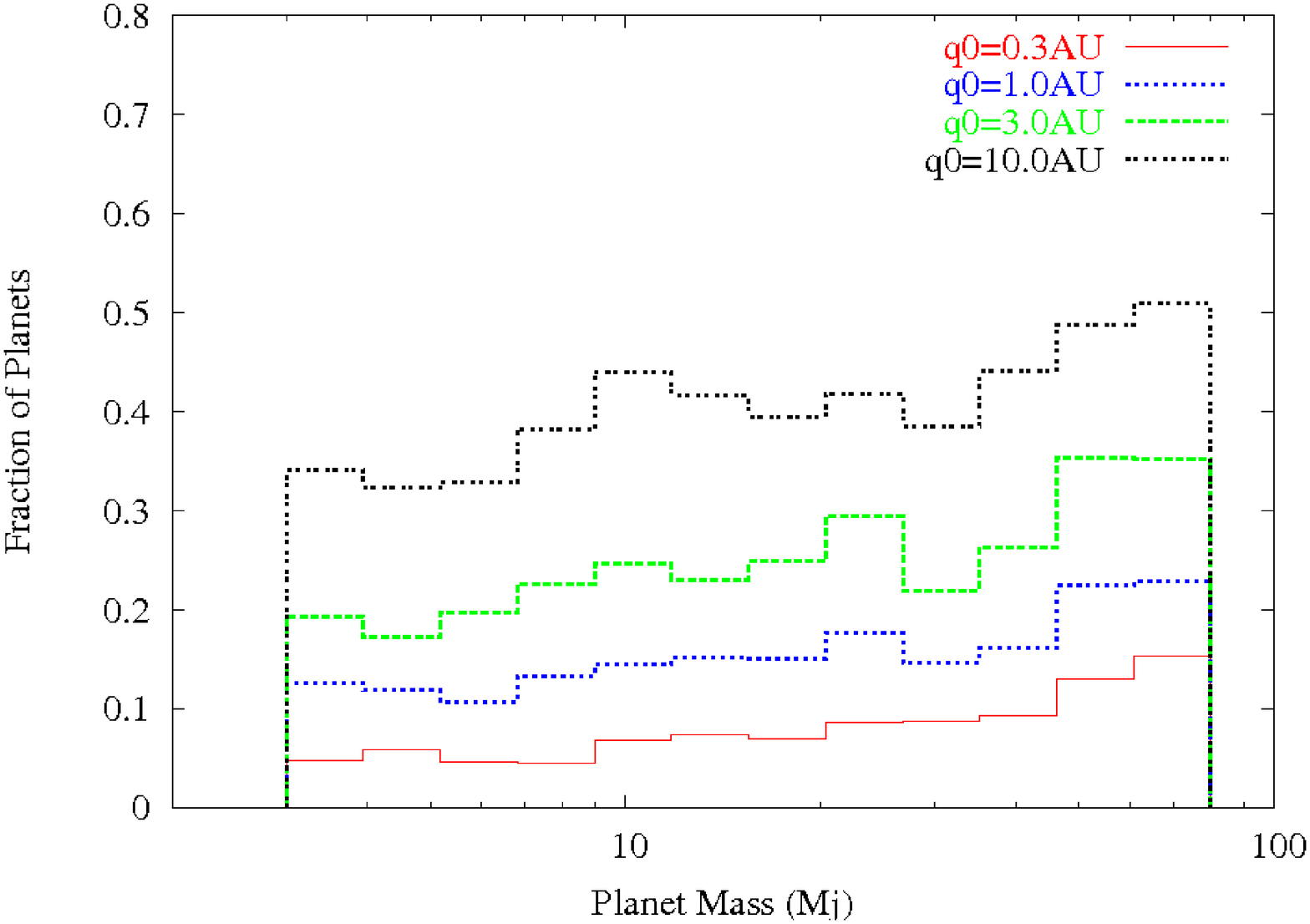}
\caption{Fraction of outer planets that spend more than 10 orbits with a 
pericenter distance below $q_{\circ}$. Two examples are given, one 
considering an initial number of two planets from each population (top) 
and another one with five planets from each population (bottom). The 
fraction of planets is relative to the initial mass distribution, which is 
constant in logarithmic scale.}
\label{nanes}
\end{figure}

\clearpage







\clearpage



\begin{table}
\centering
\begin{tabular}{ccccc}
\tableline
Initial planets & Star collisions (\%) & Planet collisions (\%) & Ejections (\%) &
Surviving \\ \tableline \tableline
2+2 & 10.9 & 1.1 & 38.0 & 2.00 \\
3+2 & 14.7 & 1.4 & 43.4 & 2.02 \\
3+3 & 14.8 & 1.7 & 51.9 & 1.90 \\
5+2 & 14.7 & 4.9 & 52.2 & 1.97 \\
4+4 & 13.4 & 1.9 & 61.9 & 1.82 \\
3+5 & 12.4 & 0.9 & 63.6 & 1.85 \\ 
5+5 & 13.8 & 2.0 & 65.3 & 1.89 \\
0+3 &  3.3 & 0.0 & 49.8 & 1.41 \\ 
0+4 &  4.0 & 0.0 & 57.0 & 1.56 \\
\tableline
\end{tabular} 
\caption{Percentage of catastrophic events for ensembles with different 
number of planets (inner+outer), and average number of surviving planets
after dynamical relaxation. Planets have four possible fates: collision
with the central star, collision with another planet, ejection, and
survival as orbiting planet. The last two rows are ensembles with the
external planet population only.\label{destins}}
\end{table}

\clearpage

\begin{table} 
\centering
\begin{tabular}{cccccc}
\tableline
Initial Planets & $a \leq 3AU (\%) $ & $a \leq10AU (\%)$ & $a \leq30AU (\%)$  \\
\tableline \tableline
2+2 & 0.8 & 2.4 & 18.0 \\
3+2 & 0.6 & 2.2 & 21.8 \\ 
3+3 & 0.4 & 4.2 & 28.6 \\
5+2 & 0.4 & 3.2 & 32.8 \\
4+4 & 0.4 & 5.6 & 39.4 \\
3+5 & 1.6 & 7.2 & 47.2 \\ 
5+5 & 1.8 & 9.6 & 58.0 \\
0+3 & 0.8 & 1.0 & 15.6 \\ 
0+4 & 0.8 & 1.4 & 25.2 \\ \tableline
\end{tabular} 
\caption{Percentage of planetary systems with at least one outer planet at 
small semimajor axis at the end of the evolution. The last two rows 
correspond to ensembles with an external population of planets only.\label{semi}} 
 \end{table}

\clearpage

\begin{table} 
\centering
\begin{tabular}{ccccc}
\tableline
$N_{planets}$ & $q_{min} \leq 0.3AU (\%)$ & $q_{min} \leq1.0AU (\%)$ & $q_{min}
\leq3.0AU (\%)$ & $q_{min} \leq10.0AU (\%)$ \\ \tableline \tableline
2+2 & 11.0 & 17.3 & 28.1 & 45.3 \\
3+2 & 11.2 & 19.1 & 30.4 & 46.1 \\ 
3+3 & 15.3 & 25.0 & 36.9 & 51.6 \\
5+2 &  6.8 & 14.8 & 26.0 & 43.5 \\
4+4 & 14.3 & 23.4 & 34.7 & 53.5 \\
3+5 & 17.1 & 26.9 & 39.0 & 55.2 \\ 
5+5 & 15.1 & 25.7 & 38.4 & 55.9 \\ 
0+3 & 19.0 & 28.7 & 39.8 & 55.5 \\
0+4 & 18.8 & 28.8 & 38.9 & 55.4 \\ \tableline
\end{tabular} 
\caption{Fraction of the outer planets that reach a minimum pericenter
over the course of their orbital evolution smaller than the quoted
values. The last two rows correspond to ensembles with 
an external population of planets only.\label{peric} }
\end{table}


\begin{thebibliography}{}

\bibitem [Alexander (2007)]{ale07} Alexander, R. 2007, arXiv:0712.0388 (to appear in New Astronomy Reviews).
\bibitem[Basri et al.(2005)]{bas05} Basri, G., Borucki, W.~J., \& Koch, D.\ 2005, New Astronomy Review, 49, 478
\bibitem[Beaulieu et al.(2006)]{bea06} Beaulieu, J.-P., et al.\ 2006, \nat, 439, 437 
\bibitem[Black(1997)]{bla97} Black, D.~C.\ 1997, \apjl, 490, L171 
\bibitem[Bodenheimer et al.(2001)]{bod01} Bodenheimer, P., Lin, D.~N.~C., \& Mardling, R.~A.\ 2001, \apj, 548, 466 
\bibitem[Bonnell et al. (2008)]{bon08} Bonnell, I., Clark, P., \& Bate, M. R. 2008, MNRAS, submitted (arXiv:0807.0460).
\bibitem[Bord{\'e} et al.(2003)]{bor03} Bord{\'e}, P., Rouan, D., \& L{\'e}ger, A.\ 2003, \aap, 405, 1137
\bibitem[Boss(1997)]{bos97} Boss, A.~P.\ 1997, Science, 276, 1836 
\bibitem[Boss(2002)]{bos02} Boss, A.~P.\ 2002, Earth and Planetary Science Letters, 202, 513 
\bibitem[Boss(2007)]{bos07} Boss, A.~P.\ 2007, \apjl, 661, L73 
\bibitem[Burrows et al.(1997)]{bur97} Burrows, A., et al.\ 1997, \apj, 491, 856
\bibitem[Butler et al. (2006)]{but06} Butler, R.~P., et al.\ 2006, \apj, 646, 505
\bibitem[Caballero et al.(2007)]{cab07} Caballero, J.~A., et al.\ 2007, \aap, 470, 903 
\bibitem[Calvet et al.(2000)]{cal00} Calvet, N., Hartmann, L., \& Strom, S.~E.\ 2000, Protostars and Planets IV, 377
\bibitem[Chabrier et al.(2005)]{cha05} Chabrier, G., Baraffe, I., Allard, F., \& Hauschildt, P.~H.\ 2005, ArXiv Astrophysics e-prints, arXiv:astro-ph/0509798 
\bibitem[Chabrier et al.(2007)]{cha07} Chabrier, G., Baraffe, I., Selsis, F., Barman, T.~S., Hennebelle, P., \& Alibert, Y.\ 2007, Protostars and Planets V, 623 
\bibitem[Chambers(1999)]{cha99} Chambers, J.~E.\ 1999, \mnras, 304, 793 
\bibitem[Chatterjee et al.(2008)]{cha08} Chatterjee, S., Ford, E.~B., Matsumura, S., \& Rasio, F.~A.\ 2008, \apj, 686, 580 
\bibitem[Close et al.(2003)]{clo03} Close, L.~M., Siegler, N., Freed, M., \& Biller, B.\ 2003, \apj, 587, 407 
\bibitem[Fischer \& Valenti(2005)]{fis05} Fischer, D.~A., \& Valenti, J.\ 2005, \apj, 622, 1102 
\bibitem[Gaudi et al.(2008)]{gau08} Gaudi, B. S., et al.\ 2008, Science, 319, 927
\bibitem[Gizis et al. (2001)]{giz01} Gizis, J. E. et al. 2001, ApJ, 551, L163
\bibitem[Gould et al.(2006)]{gou06} Gould, A., et al.\ 2006, \apjl, 644, L37 
\bibitem[Guillot et al.(2006)]{gui06} Guillot, T., Santos, N.~C., Pont, F., Iro, N., Melo, C., \& Ribas, I.\ 2006, \aap, 453, L21 
\bibitem[Haisch et al. (2001)]{hai01} Haisch, K. E., Lada, E. A., \& Lada, C. J. 2001, ApJ, 553, L153
\bibitem[Hebrard et al.(2008)]{heb08} Hebrard, G., et al.\ 2008, ArXiv e-prints, 806, arXiv:0806.0719
\bibitem[Juri{\'c} \& Tremaine(2008)]{jur08} Juri{\'c}, M., \& Tremaine, S.\ 2008, \apj, 686, 603 
\bibitem[Lin et al.(1996)]{lin96} Lin, D.~N.~C., Bodenheimer, P., \& Richardson, D.~C.\ 1996, \nat, 380, 606 
\bibitem[Mayor et al.(1998)]{may98} Mayor, M., Queloz, D., \& Udry, S.\ 1998, Brown Dwarfs and Extrasolar Planets, 134, 140 
\bibitem[McCaughrean et al.(2004)]{mcc04} McCaughrean, M.~J., Close, L.~M., Scholz, R.-D., Lenzen, R., Biller, B., Brandner, W., Hartung, M., \& Lodieu, N.\ 2004, \aap, 413, 1029 
\bibitem[Moro-Mart{\'{\i}}n et al.(2007)]{mor07} Moro-Mart{\'{\i}}n, A., et al.\ 2007, \apj, 658, 1312 
\bibitem[Murray et al.(1998)]{mur98} Murray, N., Hansen, B., Holman, M., \& Tremaine, S.\ 1998, Science, 279, 69 
\bibitem[Narita et al.(2008)]{nar08} Narita, N., Sato, B., Ohshima, O., \& Winn, J.~N.\ 2008, \pasj, 60, L1 
\bibitem[Nakajima et al.(1995)]{nak95} Nakajima, T., Oppenheimer, B.~R., Kulkarni, S.~R., Golimowski, D.~A., Matthews, K., \& Durrance, S.~T.\ 1995, \nat, 378, 463 
\bibitem[Neuh\"auser \& Guenther (2004)] {neu04} Neuh\"auser, R., \& Guenther, E. W. 2004, A\& A, 420, 647
\bibitem[Padoan \& Nordlund (1999)]{pad99} Padoan, P., \& Nordlund, A. 1999, ApJ, 526, 279
\bibitem[Padoan \& Nordlund (2004)]{pad04} Padoan, P., \& Nordlund, A. 2004, ApJ, 617, 559
\bibitem[Papaloizou \& Terquem(2001)]{pap01}Papaloizou, J.~C.~B., \& Terquem, C.\ 2001, \mnras, 325, 221 
\bibitem[Pinsonneault et al.(2001)]{pin01} Pinsonneault, M.~H., DePoy, D.~L., \& Coffee, M.\ 2001, \apjl, 556, L59 
\bibitem[Pollack et al.(1996)]{pol96} Pollack, J.~B., Hubickyj, O., Bodenheimer, P., Lissauer, J.~J., Podolak, M., \& Greenzweig, Y.\ 1996, Icarus, 124, 62 
\bibitem[Press et al. (1992)]{pre92} Press, W.H., Teukolsky, S.A., Vetterling, W.T., \& Flannery, B. P. 1992, Numerical Recipes in Fortran 77 (Cambridge University Press)
\bibitem[Queloz et al.(2000)]{que00} Queloz, D., Eggenberger, A., Mayor, M., Perrier, C., Beuzit, J.~L., Naef, D., Sivan, J.~P., \& Udry, S.\ 2000, \aap, 359, L13 \bibitem[Rafikov(2005)]{raf05} Rafikov, R.~R.\ 2005, \apjl, 
621, L69 
\bibitem[Rafikov(2006)]{raf06} Rafikov, R.~R.\ 2006, \apj, 
648, 666 
\bibitem[Rafikov(2007)]{raf07} Rafikov, R.~R.\ 2007, \apj, 
662, 642 
\bibitem[Rees(1976)]{ree76} Rees, M.~J.\ 1976, \mnras, 176, 483 
\bibitem[Ribas \& Miralda-Escud{\'e}(2007)]{rib07} Ribas, I., 
\& Miralda-Escud{\'e}, J.\ 2007, \aap, 464, 779 
\bibitem[Sicilia-Aguilar et al. (2006)]{sic06} Sicilia-Aguilar, A., \etal 2006, ApJ, 638, 897
\bibitem[Stevenson(1982)]{ste82} Stevenson, D.~J.\ 1982, \planss, 30, 755 
\bibitem[Terquem \& Papaloizou(2002)]{ter02} Terquem, C., \& 
Papaloizou, J.~C.~B.\ 2002, \mnras, 332, L39 
\bibitem[Winn et al.(2005)]{win05} Winn, J.~N., et al.\ 2005, \apj, 631, 1215 
\bibitem[Wisdom \& Holman(1991)]{wis91} Wisdom, J., \& Holman, M.\ 1991, \aj, 102, 1528
\bibitem[Zapatero Osorio et al.(2000)]{zap00} Zapatero Osorio, M.~R., B{\'e}jar, V.~J.~S., Mart{\'{\i}}n, E.~L., Rebolo, R., y Navascu{\'e}s, D.~B., Bailer-Jones, C.~A.~L., \& Mundt, R.\ 2000, Science, 290, 103 

\end{thebibliography}
\end{document}